\documentclass[twoside]{article}

\usepackage[accepted]{aistats2025}
\usepackage[round]{natbib}

\usepackage{latexsym,amsmath,amsthm,amsfonts,amssymb,mathrsfs,graphicx,hyperref,mathpazo,soul,color,comment}
\usepackage{caption}
\usepackage{subcaption}
\usepackage{relsize}
\usepackage{bm}

\usepackage{hhline}
\usepackage{multirow}

\usepackage{algorithm}
\usepackage{algorithmic}
\usepackage{etoolbox}
\AtBeginEnvironment{algorithm}{%
  \setlength{\columnwidth}{\linewidth}%
}
\usepackage{etoolbox}

\usepackage{array}

\def\erf{\mathop{{\rm erf}}}
\def\argmin{\mathop{{\rm argmin}}}

\newcounter{problem}
\newenvironment{problem}[1][]{\refstepcounter{problem}\par\medskip
   \noindent \textbf{Problem~\theproblem.  #1 }  \rmfamily}{\medskip}
\theoremstyle{plain}

\newtheorem{theorem}{Theorem}[section]

\theoremstyle{definition}

\newtheorem{example}{Example}

\theoremstyle{remark}

\begin{document}

%

%

\twocolumn[

\aistatstitle{Nonparametric Distributional Regression via  Quantile Regression }

\aistatsauthor{ Cheng Peng \And Stan Uryasev}

\aistatsaddress{ Stony Brook University} ]

\begin{abstract}
  This paper proposes a new approach to estimating the distribution of a response variable conditioned on factors. We model the conditional quantile function as a mixture (weighted sum) of basis quantile functions, with weights depending on these factors. The estimation problem is formulated as a convex optimization problem. The objective function is equivalent to the continuous ranked probability score (CRPS). This approach can be viewed as conducting quantile regressions for all confidence levels simultaneously while inherently avoiding quantile crossing. We use spline functions of factors as a primary example for the weight function. We prove an approximation property of the model. 
  To address computational challenges, we propose a dimensionality reduction method using tensor decomposition and an alternating algorithm. Our approach offers flexibility, interpretability, tractability, and extendability. Numerical experiments demonstrate its effectiveness.
\end{abstract}

\section{INTRODUCTION}

This paper studies 
estimating the distribution of  the response variable  $y$ conditioned on observing some factors (features) $\bm{x}$ given data pairs $\{(y_i,\bm{x}_i)\}_{i=1}^N$. 
The quantile function of random variable $y$ is defined by
$
Q_y(p) = \inf_q\{ p \leq P(y\leq q) \}
$. 

Our main contributions are as follows:
\begin{itemize}
\item We develop a general model of conditional quantile function (Section \ref{sec_model}). The conditional quantile function $Q_{y|\bm{x}}(p)$ 
is modeled by a mixture (weighted sum) of  basis quantile functions. 
The weight of each  basis quantile function is a function of the factors. We  propose various possible choices of basis functions for weight and quantile.
\item  We formulate the parameter estimation problem as a convex optimization (Section \ref{sec_calibration}). The formulation takes advantage of the linearity of the model with respect to the parameters. 
\item The parameter estimation problem can be viewed as constrained multiple quantile regression \citep{koenker2005book} (Section \ref{sec_quantile_reg}). The model formulation rules out quantile crossing (Section \ref{sec_noncrossing}).
\item  We study the approximation property of the model (Section \ref{sec_approximation}). 
\item  We  propose a dimensionality reduction method using tensor decomposition(Section \ref{sec_dim_reduction}). The number of parameters grows linearly w.r.t. the dimension of factor. We propose an alternating algorithm for computation.
\end{itemize}

While the quantile functions and weight functions are arbitrary in principle, we primarily focus on quantile functions of common distributions and B-splines  with nonnegative coefficients \citep{Boor1974nonnegativeB}.  

 The model possesses an appealing interpretation that the factors have different impact on the body and the tail of the conditional distribution. The impact of  factors on the shape of conditional quantile function can be traced analytically.  Both the conditional quantile function and the quantile  (hyper)surface have closed-form expressions (Section \ref{sec_model}). 



\section{LITERATURE REVIEW}\label{sec_review}


\textit{Mixture quantiles model} \citep{tukey1962future, gilchrist2000statistical} uses a combination of some basis functions  functions to model the quantile function. Various basis functions have been used  \citep{Sillitto1969, Karvanen2006mixtures, Keelin2016metalog}.

 \textit{Mixture regression} models the conditional distribution as a linear combination of basis functions  \citep{Quandt1972switching,
desarbo1988regression,
VILLANI2009155mixture,
Yuksel2012reviewmixture}.

\textit{Quantile regression} 
estimates the conditional quantile of a given confidence level given  some factors. 
Spline  quantile regression has been presented in  \cite{koenker1994spline,He1998bispline,
Koenker2011additive}. Quantile crossing are addressed in \cite{bondell2010noncrossing,
 Chernozhukov2010rearrange}. 
 
 \textit{Quantile regression process} \citep{Angrist2006qrpasymptotics,Reich2011Bqrprocess,
LIAN2015qrprocess, Frumento2016WQR, Yuan2017Splineqrprocess} refers to the regression coefficient as a function of the confidence level of quantile regression. Such model does not prevent quantile crossing. .

  \textit{Continuous Ranked Probablity Score} (CRPS, \cite{Matheson1976score}) is a proper scoring rule  that is frequently used to measure the quality of a distributional prediction and as the objective function of optimization \citep{Hothorn2014Transformation,duan2020ngboost,
Gouttes2021NN,BERRISCH2021CRPS}. 
 \cite{Laio2006VerificationTF} proposes an equivalent definition that is useful to formulate convex optimization.

There is a variety of distributional regression models. To list a few, \textit{Conditional transformation model} \citep{Hothorn2014Transformation},  \textit{generalized additive models for location, scale and shape} (GAMLSS,  \cite{Rigby2005gamlss}),
 \textit{The linear and polynomial model} in \cite{chernozhukov2001conditional}, 
\textit{Spline quantile function RNN}  \citep{Gasthaus2019quantile}.

\textit{Canonical polyadic (CP) decomposition} \citep{Hitchcock1927cp,
Harshman1970cp,Carroll1970cp} for spline tensors has been applied in Computer Aided Design (CAD) to reduce model complexity \citep{Pan2016cad}. 

\paragraph{Comparision}
 Compared to existing literature, our approach has the following unique properties: \textit{(i)} modeling the entire conditional quantile function with good approximation properties, closed-form expression and inherent prevention of quantile crossing, \textit{(ii)} convex optimization to find the global optimum in parameter estimation, \textit{(iii)} providing interpretability by allowing observation of how changes in factors influence the distribution's shape, and \textit{(iv)} dimensionality reduction using tensor decomposition. A detailed literature review is in Appendix \ref{sec_review_append}.

\section{MODEL DESCRIPTION}\label{sec_model}

This section describes the Factor Model of Mixture Quantiles and discusses noncrossing quantile  and various perspectives to view the model. 

\noindent
\begin{itemize}
\item $p = $ confidence level of a quantile function
\item $\bm{x} = $ vector of factors
\item $\bm{a}=$  vector of parameters (unknown coefficients to be estimated) 
\item $Q(p)=$ quantile function
\item $G(p,\bm{x},\bm{a})=$  model with  parameters $\bm{a}$ that outputs  the $p$-quantile of the response variable conditioned on observing factor $\bm{x}$
\item $\emph{I} = $ number of basis functions in the  mixture
\item $\emph{J}  =$ number of basis functions
\end{itemize}

\subsection{Factor Model of Mixture Quantiles}

The Factor Model of Mixture Quantiles is defined as
\begin{equation} \label{model}
G(p,\bm{x},\bm{a})
=
\sum_{i=0}^\emph{I}     f_i (\bm{x},\bm{a}_i)   Q_i(p)  \;,
\end{equation}
where $Q_{0}(p) = 1$. 
The  basis functions $\{Q_i(p)\}_{i=0}^\emph{I}$ are defined on the unit interval, and are linearly independent, i.e., any $Q_i(p)$ is not equal to a linear combination of  other $Q_j(p)$, $j\neq i$. The weight function $f_i (\bm{x},\bm{a}_i)$ of each basis function $Q_i(p)$ is a function of the factors $\bm{x}$.

The model formulation \eqref{model} is quite general since $\{Q_i(p)\}_{i=1}^\emph{I}$ and  $\{ f_i (\bm{x},\bm{a}_i)  \}_{i=0}^\emph{I}$ can be arbitrary functions, providing great flexibility in modeling heteroskedastic data and nonlinear relations. The weight functions determine how the factors impact the scale of basis quantile functions. It has the appealing interpretation that different factors may impact the different part of the distribution. The conditional quantile function  has a closed-form expression if all basis quantile functions do.  With our model, Monte Carlo simulation can be easily conducted with inverse transform sampling.

We focus on the case where  $\{ f_i (\bm{x},\bm{a}_i)  \}_{i=0}^\emph{I}$ are splines with basis spline functions $\{B_{ij}(\bm{x})\}_{i,j=0}^{\emph{I},\emph{J}}$. 
Furthermore, we use the same basis spline functions for all  basis quantile functions $\{Q_i(p)\}_{i=1}^\emph{I}$, i.e., $\forall i,j, \;B_{ij}(\bm{x}) = B_{j}(\bm{x})$.  Then the model is defined as 
\begin{align}\label{model_2}
G(p,\bm{x},\bm{a})
=
\sum_{i=0}^\emph{I}  \sum_{j=0}^\emph{J} 
a_{i j }
B_{j}(\bm{x}) 
Q_i(p) 
\;,
\end{align}
where $B_0(x) = 1$, $Q_0(p)=1$. 
 Spline is adaptive to the data and can be optimized with the model in one shot. The linearity with respect to the coefficients not only results in a convenient formulation of convex optimization, but also retains an interpretable factor model structure such that we can analyze the impact of each factor on the shape of the conditional quantile function.
 
 To distinguish $\{Q_i(p)\}_{i=0}^\emph{I}$ and $\{B_j(\bm{x})\}_{j=0}^\emph{J}$, we hereafter refer to $\{Q_i(p)\}_{i=0}^\emph{I}$ as basis quantile functions and $\{B_j(\bm{x})\}_{j=0}^\emph{J}$ as  basis spline functions. The meaning of the terms may be clearer if we expand the summation
 \begin{equation}\label{model_expand}
\begin{aligned}
G(p,\bm{x},\bm{a})
=
& a_{00}
+
 \sum_{j=1}^\emph{J} 
a_{0 j }
B_{j}(\bm{x}) 
+
\sum_{i=1}^\emph{I}   
a_{i 0 } 
Q_i(p) 
\\
& +
\sum_{i=1}^\emph{I}    \sum_{j=1}^\emph{J} 
a_{i j }
B_{j}(\bm{x}) 
Q_i(p)  \;,
\end{aligned}
\end{equation}
where $a_{00}$ is the constant location parameter, $ \sum_{j=1}^\emph{J} a_{0 j } B_{j}(\bm{x})$ determines the conditional location, $\sum_{i=1}^\emph{I}  a_{i 0 } Q_i(p)$ is the base quantile function that does not vary with factors, $\sum_{i=1}^\emph{I}  \sum_{j=1}^\emph{J}  a_{i j } B_{j}(\bm{x}) Q_i(p)$ determines the conditional quantile function.

  The complexity of the model is determined by the number of basis quantile functions, the number of factors and the form of weight functions.  
 The number of parameters in the model is $(\emph{I}+1) \times (\emph{J}+1)$. 
The degrees of freedom are usually smaller due to constraints in optimization.

 \subsection{Noncrossing Quantile Model}\label{sec_noncrossing}

 This section studies general conditions for $\{ B_j (\bm{x})  \}_{j=1}^\emph{J}$ and  $\{Q_i(p)\}_{i=1}^\emph{I}$ to guarantee noncrossing quantiles. 
 We refer to a quantile model as  noncrossing if it satisfies that the conditional quantile functions of any two different confidence levels do not cross  conditioned on any value of factors. That is, 
 \begin{equation}\label{noncrossing}
 \begin{aligned}
 \forall \bm{x} \in &\mathcal{X},p_1,p_2,\; 0< p_1 < p_2<1,\;  \\
 &G(p_1,\bm{x},\bm{a}) \leq  G(p_2,\bm{x},\bm{a}) \;,
 \end{aligned}
 \end{equation}
 where $\mathcal{X}$ is the set of all possible values of factors $\bm{x}$. In the literature, the condition is often relaxed to that $ G(p_1,\bm{x},\bm{a}) \leq  G(p_2,\bm{x},\bm{a})$ for a certain $\bm{x} \in \mathcal{X}$ or for a certain selected confidence levels $\{p_m\}_{m=1}^\emph{M}$.  
  With such simplification, it could still require a large number of constraints to guarantee noncrossing. 
  
\paragraph{Sufficient condition} Instead of relaxing the condition  \eqref{noncrossing}, we propose using a sufficient condition.  Note that the model \eqref{model_2} satisfies noncrossing condition \eqref{noncrossing} if three  conditions are satisfied: \textit{(i)} $\{Q_i(p)\}_{i=1}^\emph{I}$ are nondecreasing;
\textit{(ii)} $\{ B_j (\bm{x})  \}_{j=1}^\emph{J}$ are nonnegative;
\textit{(iii)} $\{a_{ij}\}_{i\neq 0}$ are nonnegative.

  Nonnegativity constraint is relatively easy to impose in optimization.  Although a sufficient condition may seem too restrictive,  Theorem \ref{theorem_approx} in Section \ref{sec_approximation} shows that the model retains good approximation ability.

\paragraph{Basis quantile function} For  $\{Q_i(p)\}_{i=1}^\emph{I}$, the following two types of nondecreasing functions can be used:
\textit{(i)} quantile functions of common distributions;
\textit{(ii)}  monotone basis spline functions such as I-spline \citep{Ramsay1988Ispline}.

 Quantile functions of common distributions are prefered when the shape of the distribution is known to be close to common distributions. When the distribution is multimodal, splines offer greater flexibility. Since splines are bounded on the unit interval, it needs to be combined with common quantile functions to model distributions with long tails. 
   
\paragraph{Basis weight function} For $\{ B_j (\bm{x})  \}_{j=1}^\emph{J}$, the following three types of nonnegative function can be used: \textit{(i)}
 nonnegative basis spline functions such as B-spline  with nonnegative coefficients \citep{Boor1974nonnegativeB,
Papp2011polynomial, Papp2014spline}; 
\textit{(ii)}  arbitrary basis spline functions with constraints on coefficients that characterize nonnegative polynomials 
\citep{lukacs1918polynomial,
Papp2011polynomial, Papp2014spline};
\textit{(iii)} nonnegative kernel functions \citep{marteau2020nonnegative}.

While the second type is more flexible than the first one, it is more computationally expensive to optimize. 
   Besides B-spline, other choices include Bernstein polynomial  and M-spline \citep{Curry1988Mspline}. B-spline and M-spline differ by a constant. \cite{Papp2011polynomial}, \cite{Papp2014spline} show that piecewise Bernstein polynomial includes B-spline as a subset. 
  The monotone I-spline used for $\{Q_i(p)\}_{i=1}^\emph{I}$ is obtained by integrating M-spline. 


\subsection{Examples}

This subsection provides several  examples of the model described in Section \ref{sec_model}.  We refer to  models from other research to  demonstrate the wide applicability of our approach. 

\begin{example}\label{eg_locationscale}
Location-scale model of normal distribution
\begin{align*}
G(p,\bm{x},\bm{a})
=
a_{00}  +  \sqrt{2}{\erf}^{-1}(2p-1) a_{10}   \;,
\end{align*}
where  $\sqrt{2}{\erf}^{-1}(2p-1)$ is the quantile function of standard normal distribution.
\end{example}

\begin{example} \label{eg_mixture}
Logistic-normal  mixture of quantiles without factor dependence \citep{Karvanen2006mixtures,
Keelin2016metalog,
Cheng2022mixquantile}
\begin{equation*}
\begin{aligned}
G(p,\bm{x},\bm{a})
\!=\!
a_{00} \! +\!   \log(\frac{p}{1-p})  a_{10}
 \! + \! \sqrt{2}{\erf}^{-1}(2p-1) a_{20} ,
 \end{aligned}
\end{equation*} 
where  $\log(\frac{p}{1-p})$ is the quantile function of standard logistic distribution. 
\end{example}

\begin{example}\label{eg_linearnormal}
Linear model with normal noise
\begin{equation*}
G(p,\bm{x},\bm{a}) = a_{00} + x a_{01}  + \sqrt{2}{\erf}^{-1}(2p-1) \;.
\end{equation*}
\end{example}

\begin{example}\label{eg_qrproess}
Quantile regression process model \citep{Reich2011Bqrprocess,
LIAN2015qrprocess,
Frumento2016WQR,
Yuan2017Splineqrprocess}
\begin{align*}
G(p,\bm{x},\bm{a})
=
 \sum_{k_1=1}^{K} B_{k_1}(p) x_1 a_{1 k_1} +
 \sum_{k_2=1}^{K} B_{k_2}(p) x_2  a_{2 k_2} \;,
\end{align*}
where $B_{k}(p)$ can be  Bernstein basis polynomials or B-spline basis. 
\end{example}

\begin{example}\label{eg_arch}
Autoregressive conditional heteroskedasticity-1  (ARCH(1)) model with normal noise \citep{Engle1982arch}
\begin{equation*}
G_t(p,\bm{x},\bm{a})
=
 \sqrt{2}{\erf}^{-1}(2p-1)  \cdot \sqrt{ a_{10} +  x_{t-1}^2 a_{11} }  \;,
\end{equation*}
where $G_t$ is the quantile function of $x_t$ at time $t$, $x_{t-1}$ is the response variable at at time $t-1$. At every time $t$, the quantile function $G_t(p,\bm{x},\bm{a})$  depends on  the previous  response variable $x_{t-1}$.
\end{example}

\begin{example} \label{eg_dynamic}
Dynamic quantile model \citep{
Gourieroux2008dynamic}
\begin{equation*}
\begin{aligned}
G_t(p,\bm{x},\bm{a})
\!=\!
 a_{00} +  x_{t-1} a_{01} 
\!+ \! ( a_{10} \!+ \! x_{t-1}^2 a_{11})\log(\frac{p}{1-p}) .
\end{aligned}
\end{equation*}
where $G_t$ is the quantile function of $x_t$ at time $t$, $x_{t-1}$ is the response variable at at time $t-1$. At every time $t$, the quantile function $G_t(p,\bm{x},\bm{a})$  depends on  past response variable $x_{t-1}$. 
 The response variable at time $t$ is generated by plugging a sample $p$ from uniform distribution $U(0,1)$ and the past response variable $x_{t-1}$ to $G_t(p,\bm{x},\bm{a})$.
\end{example}

 Example \ref{eg_locationscale}, \ref{eg_linearnormal} and \ref{eg_arch}  shows that our model incorporates some of the most common models as special cases. Maximum likelihood estimation for these models are well studied. We propose a different estimation method in Section \ref{sec_calibration}. Dynamic models such as Example \ref{eg_arch} and \ref{eg_dynamic} provide useful complements to classic time series models.  
Although we focus on linear function in $\bm{a}$ in subsequent sections on parameter estimation, the function   $f(\bm{x},\bm{a})$ in Example \ref{eg_arch}  is  nonlinear in both $\bm{x}$ and $\bm{a}$. The parameters can still be estimated by optimizing our proposed objective function, but the optimization is not convex.   
\cite{
Gourieroux2008dynamic} also uses $|x|$ and $x^+ = \max\{0,x\}$, $x^- = \max\{0,-x\}$ to replace $x$, so that the factors remain nonnegative. 

\section{PARAMETER ESTIMATION BY CONVEX OPTIMIZATION}
\label{sec_calibration}

This section includes  the basics of quantile regression \citep{KoenkerBassett1978quantile}  and the convex formulation for parameter estimation problem.

\subsection{Quantile Regression}\label{sec_quantile_reg}

Consider a model of $p$-quantile  of response variable $y|\bm{x}$ conditioned on factor $\bm{x}$ with parameter $\bm{a}$
\begin{align}
q_p(y|\bm{x}) = G(p,\bm{x},\bm{a}) \;.
\end{align}
Quantile regression estimates the parameter $\bm{a}$ by minimizing the  scaled Koenker-Bassett error of the residuals
\begin{align}
\min_{\bm{a}} \bm{\mathcal{E}}_p(y-G(p,\bm{x},\bm{a}))
\end{align}
where the error is the expected pinball loss 
\begin{align}
\bm{\mathcal{E}}_p(Z) &= E[\rho_p(Z)] \;, \\
\rho_p(Z) &= pZ \mathbb{1}_{\{ Z > 0 \}} - (1-p) Z \mathbb{1}_{\{ Z \leq 0 \}} \;, \label{def_pinball}
\end{align}
$Z$ is a  random variable, $\mathbb{1}_{\{\cdot\}}$ is the indicator function that equals $1$ when the equation in the bracket is true and $0$ if otherwise.  

\subsection{Optimization Problem Statement for Parameter Estimation}

\begin{itemize}
\item $\emph{M}=$ number of grid points of discretized confidence level
\item $\emph{N}=$ sample size
\item $\bm{Y}^\emph{N} = (y_1,\cdots,y_\emph{N})=$ response variables of dependent variable
\item $\bm{x}_n=$ vector of factors  corresponding to response variable $y_n$, $n=1,\cdots,\emph{N}$;
\item $\bm{x}^\emph{N}=(\bm{x}_1,\cdots,\bm{x}_\emph{N})= $  set of factor vectors
\item $\bm{a}=$  vector of parameters
\item $\mathcal{A}=$  feasible set of $\bm{a}$ determined by constraints
\item $L(\bm{a};\bm{Y}^\emph{N},\bm{x}^\emph{N},p)=$  discrete residual random variable taking with equal probabilities the following values 
$$y_n -   
\sum_{i=0}^\emph{I}  \sum_{j=0}^\emph{J} 
a_{i j }
B_{j}(\bm{x}_n) 
Q_i(p)   
\;,
\quad n=1,\ldots,\emph{N} 
$$ 
\end{itemize}

\bigskip

The optimization problem statement for finding optimal parameters $\bm{a}$ is formulated as follows
\begin{problem}\label{problem_1}
\begin{equation}
	\label{continuous}
	\begin{aligned}
		\min_{\bm{a}\in \mathcal{A}} \; &   \int_0^1  w(p) \bm{\mathcal{E}}_{p} \left( L(\bm{a};\bm{Y}^\emph{N},\bm{x}^\emph{N},p) \right) \mathrm{d}p 
		\;,
	\end{aligned}
\end{equation}
where $w(p)$ is a nonnegative weight function satisfying $\int_0^1 w(p) \mathrm{d}p= 1$. 
\end{problem}

$w(p)$ can be chosen to focus on the distribution tail or body. Inherent from the pinball loss, the estimator is robust to outliers. 
While we consider the case where $f_{i}(\bm{x},\bm{a})$ are spline functions, the estimation method works for  the general case \eqref{model}.

Next, we discretize the problem \eqref{continuous} by using a grid on $p$. The resultant  optimization problem is still a convex programming problem.

\begin{problem}
\label{first_method}
Discrete variant:
\begin{equation}\label{problem_statement}
\begin{aligned}
\min_{\bm{a} \in \mathcal{A}} \; &   \sum_{m=1}^\emph{M}  w_m \bm{\mathcal{E}}_{p_m} \left( L(\bm{a};\bm{Y}^\emph{N},\bm{x}^\emph{N},p_m) \right) 
 \;,
\end{aligned}
\end{equation}
where  $w_m$ are nonnegative weights satisfying  $\sum_{m=1}^\emph{M} w_m = 1$.
\end{problem} 

Similar to quantile regression, the problem can be reduced to linear programming. Note that although we select only a finite number of confidence levels in the discrete  variant, the model still produces the entire conditional quantile function and ensures noncrossing quantiles. The estimation method also allows one to focus on the tail of the distribution by assigning higher weights on errors with tail confidence levels.  

\subsection{Equivalence to Constrained Joint Quantile Regression}\label{two_step_procedure}

Define $\emph{M}$ discrete random variables $\widehat{L}_m(\bm{a};\bm{Y}^\emph{N},\bm{x}^\emph{N})$,  each taking with equal probabilities the values 
$ y_n -   
   \sum_{j=0}^\emph{J} 
\lambda_{ jm }
B_{j}(\bm{x}_n)  
, \quad
 n=1,\ldots,\emph{N}
$, where $\lambda$ is defined below. 
We can write \eqref{problem_statement} equivalently as
\begin{problem} \label{equivalent_formulation_0}
\begin{align}
\min_{\bm{a}\in \mathcal{A},\, \bm{\lambda} }\; &  
 \sum_{m=1}^\emph{M}  w_m \bm{\mathcal{E}}_{p_m} \left( \widehat{L}_m(\bm{a};\bm{Y}^\emph{N},\bm{x}^\emph{N})\right) \label{objective_0}
\\
\text{subject to} \; 
&  \sum_{i=0}^\emph{I} 
 Q_i(p_m) a_{ij} 
 = \lambda_{jm}  \;, \label{sys_equations_0}
\\
   & m = 1,\ldots,\emph{M},\; j=0,\cdots,\emph{J}
 \;.  \nonumber
\end{align}
\end{problem}
This  problem formulation makes it clear that the estimation in Problem \ref{first_method}  is equivalent to conducting several quantile regressions in one shot with constraints on the parameters. Since  $\sum_{j=0}^\emph{J} 
\lambda_{ jm }
B_{j}(\bm{x}_n) $ is a spline of scalar factor $x_j$, the objective function in \eqref{objective_0} is the sum of objective functions of  spline quantile regressions with confidence levels $\{p_m\}_{m=1}^\emph{M}$. 
$\{\lambda_{jm}\}_{j=0,m=1}^{\emph{J},\emph{M}}$ are associated by systems of equations \eqref{sys_equations_0}, where the coefficients $\bm{a}$ are constrained by feasible set $\mathcal{A}$.

  As $\emph{J}$ systems of equations, \eqref{sys_equations_0} can be written in matrix format
	$\bm{Q} \bm{A} = \bm{\Lambda}$, 
where $\bm{Q} = [Q_i(p_m)]_{\emph{M} \times (\emph{I}+1)}',
 \bm{A} = 
 [a_{ij}]_{(I+1) \times (\emph{J}+1)},
 \bm{\Lambda} = 
 [\lambda_{jm}]_{\emph{M} \times (\emph{J}+1)}'$.

 If the solution to the unconstrained problem \eqref{objective_0} is in the feasible set defined by $\mathcal{A}$ and \eqref{sys_equations_0}, we can solve Problem \ref{equivalent_formulation_0} with the following two steps.
\begin{enumerate}
\item  Solve $\emph{M}$ spline quantile regressions separately
\begin{align}\label{problem_statement_2}
  \!\! \!\! \min_{ \bm{\lambda} }\;  &   \bm{\mathcal{E}}_{p_m} \left(\widehat{L}_m(\bm{a};\bm{Y}^\emph{N},\bm{x}^\emph{N})  \right), \;m=1,\cdots,M .
\end{align}
\item  Solve the systems of equations with constraints
\begin{equation}\label{second_step}
	\begin{aligned}
		 \bm{a}& \in\mathcal{A} ,\;\; 
		 \bm{Q} \bm{A} &= \bm{\Lambda}  \;.
	\end{aligned}
\end{equation}
\end{enumerate}

In both steps, the solution can be nonunique. The uniqueness of solution to quantile regression is discussed in \cite{koenker2005book}. 
\cite{Portnoy1991qrbreakpoints} shows that the number of distinct solutions to quantile regression when the confidence level $p$ varies in $(0,1)$ is $\mathcal{O}(N\log N)$ in probability, which grows slower than the upper bound $\binom{N}{M \times (\emph{J}+1)}$. The uniqueness of solution to the systems of equations  depends on  $\bm{Q}$ when $\mathcal{A} = \mathrm{R}^{(I+1) \times (\emph{J}+1)}$.  For proper choice of independent basis quantile functions $\{Q_i(p)\}_{i=1}^\emph{I}$, $\bm{Q}$  has full column rank when $\emph{M}\geq \emph{I}+1$. 
To check feasibility when $\mathcal{A} = \mathrm{R}_{\geq 0}^{(I+1) \times (\emph{J}+1)}$, we can solve the linear programming problem $\min_{\bm{a} \geq 0} ||  \bm{Q} \bm{A} -  \bm{\Lambda}  ||_1$.

\subsection{Equivalence to CRPS Minimization}\label{sec_crps}

Continuous Ranking Probability Score (CRPS) is frequently used to evaluate the quality of a distributional forecast. 
We show that Problem \ref{problem_1} is equivalent to CRPS minimization. 

For a CDF $F$ and an observation $y_i$ of the response variable $y$, CRPS is defined by 
\begin{align}\label{crps_cdf_def}
CRPS(F,y_i) = \int_\mathcal{R} \left( F(x) - 1_{\{y_i \leq x \} } \right)^2 \mathrm{d}x \;.
\end{align}
This is the squared distance between $F$ and the CDF of a single observation $y_i$ of the response variable $y$. 
For the corresponding quantile function $Q$, i.e., the generalized inverse function of $F$, and the response variable $y_i$, CRPS has the following equivalent definition \citep{Laio2006VerificationTF,
Fakoor2021Aggregation}
\begin{align}
CRPS(Q,y_i) = 2\int_0^1 \rho_p \left( y_i-Q(p) \right)\mathrm{d}p \;.
\end{align}

Consider estimating the parameters by  minimizing  the sum of CRPS of the response variables 
\begin{align}\label{crps_min}
\min_{\bm{a}}\; \sum_{n=1}^\emph{N} \int_0^1 \rho_p \left( y_n-G(p,\bm{x}_n,\bm{a}) \right)\mathrm{d}p \;.
\end{align}
 We see that \eqref{crps_min} is equal to the objective function in Problem \ref{problem_1} with uniform weight $w(p)$ by exchanging the integral and sum.

 Consider the special case where the basis quantile functions $\{Q_i(p)\}_{i=0}^\emph{I}$ is $Q_0(p) = 1$. The model  $G(p,\bm{x},\bm{a})$ is not dependent on $p$ and reduces to point estimation given $\bm{x}$. Since $F$ becomes a step function, CRPS \eqref{crps_cdf_def} reduces to 
$
CRPS(F,y_i) = |y - y_i|
$, where $y$ is the discontinuous point of $F$.  Thus CRPS minimization reduces to the least absolute deviation regression. 

\section{APPROXIMATION THEOREM}\label{sec_approximation}

Our model can approximate any bounded conditional quantile model to arbitrary precision as the number of knots tends to infinity. Since the considered model has nonnegative parameters, classic approximation theorem of polynomials cannot be directly applied. The proof is in Appendix \ref{sec_proof}.

We adopt the following notations.
 $\mathcal{X}=$ hyperrectangular domain of factors.
$T= (t_{ij})_{i=1,\cdots,\emph{I+1}, j=1,\cdots,\emph{J} }=$ matrix representation of  subdivision of $[0,1]  \times  \mathcal{X} $.
 $||T||=\max_{i,j}{|t_{i+1,j}-t_{i,j}|}$ mesh size of subdivision $Z$.
$\mathrm{cone}(U)=$  cone of nonnegative linear combination of functions in a set of basis functions $U$.
 $\mathrm{int}=$ interior of a set.
 $\mathcal{P}(U,T)=$ set of piecewise functions where each piece (in
the scaled representation) defined on a subdivision defined by $Z$ is in $\mathrm{cone}(T)$.
$G_{[0,1]\times \mathcal{X}}=$ cone of all continuous functions defined on $[0,1]\times \mathcal{X}$ that is nondecreasing in the first variable.

\begin{theorem}\label{theorem_approx}
Consider I-spline basis $\{Q_i(p)\}_{i=0}^\emph{I}$ defined on $[0,1]$ and B-spline basis $\{B_j(\bm{x})\}_{j=0}^\emph{J}$ defined on $\mathcal{X}$.
 Furthermore, let $T_i$ be an asymptotically
nested sequence of subdivisions with mesh sizes approaching zero. Then the set $\cup_i \mathcal{P}(Q_i(p)  \otimes B_j(\bm{x}), T_i)$ is a
dense subcone of $G_{[0,1]\times \mathcal{X}}$.
\end{theorem}



\begin{figure*}[t]
    \centering
    \begin{minipage}{0.32\linewidth}
        \centering
        \includegraphics[width=\linewidth]{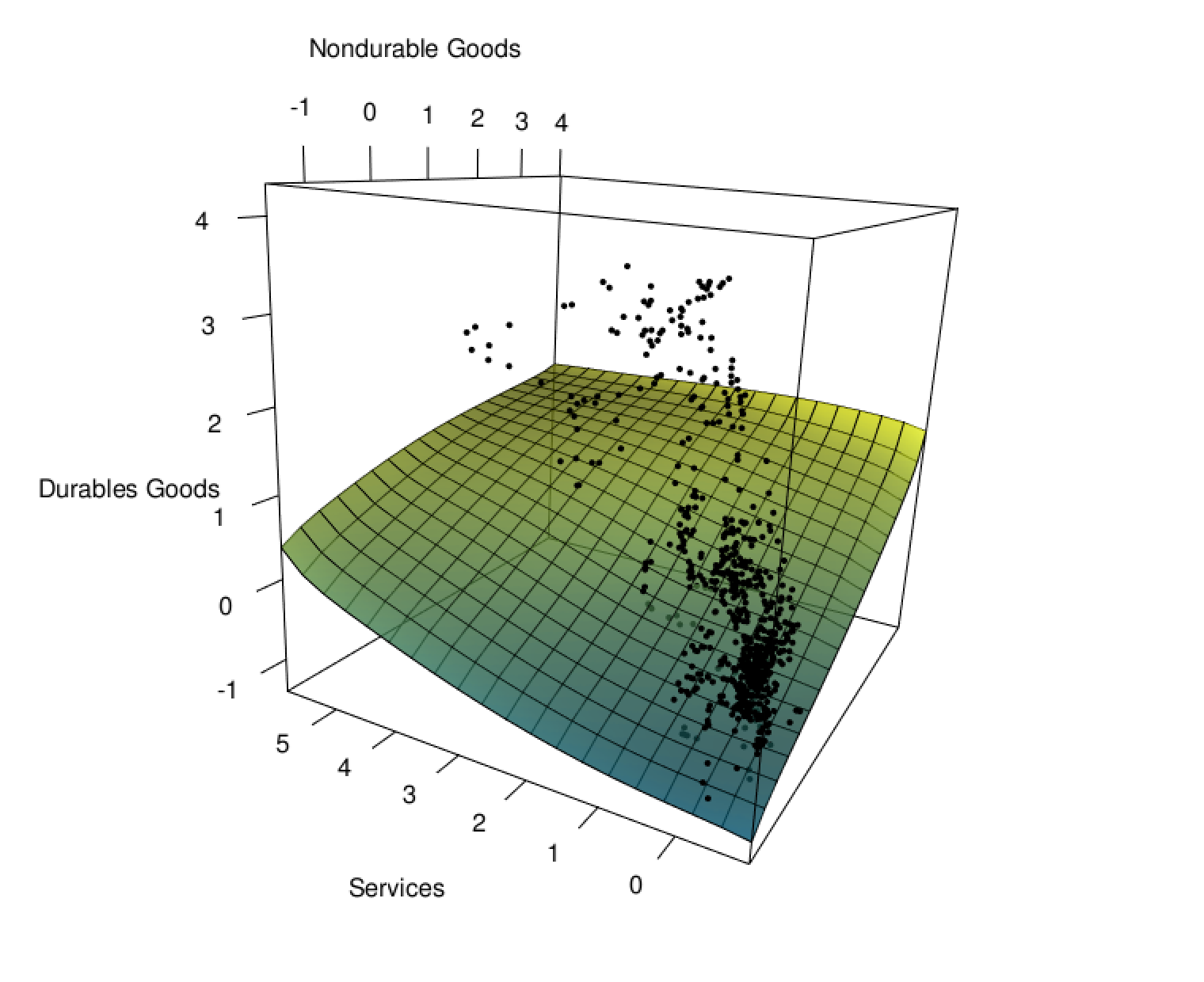}
        \caption*{Quantile surface, confidence level $p=0.05$}
    \end{minipage}
    \hfill
    \begin{minipage}{0.32\linewidth}
        \centering
        \includegraphics[width=\linewidth]{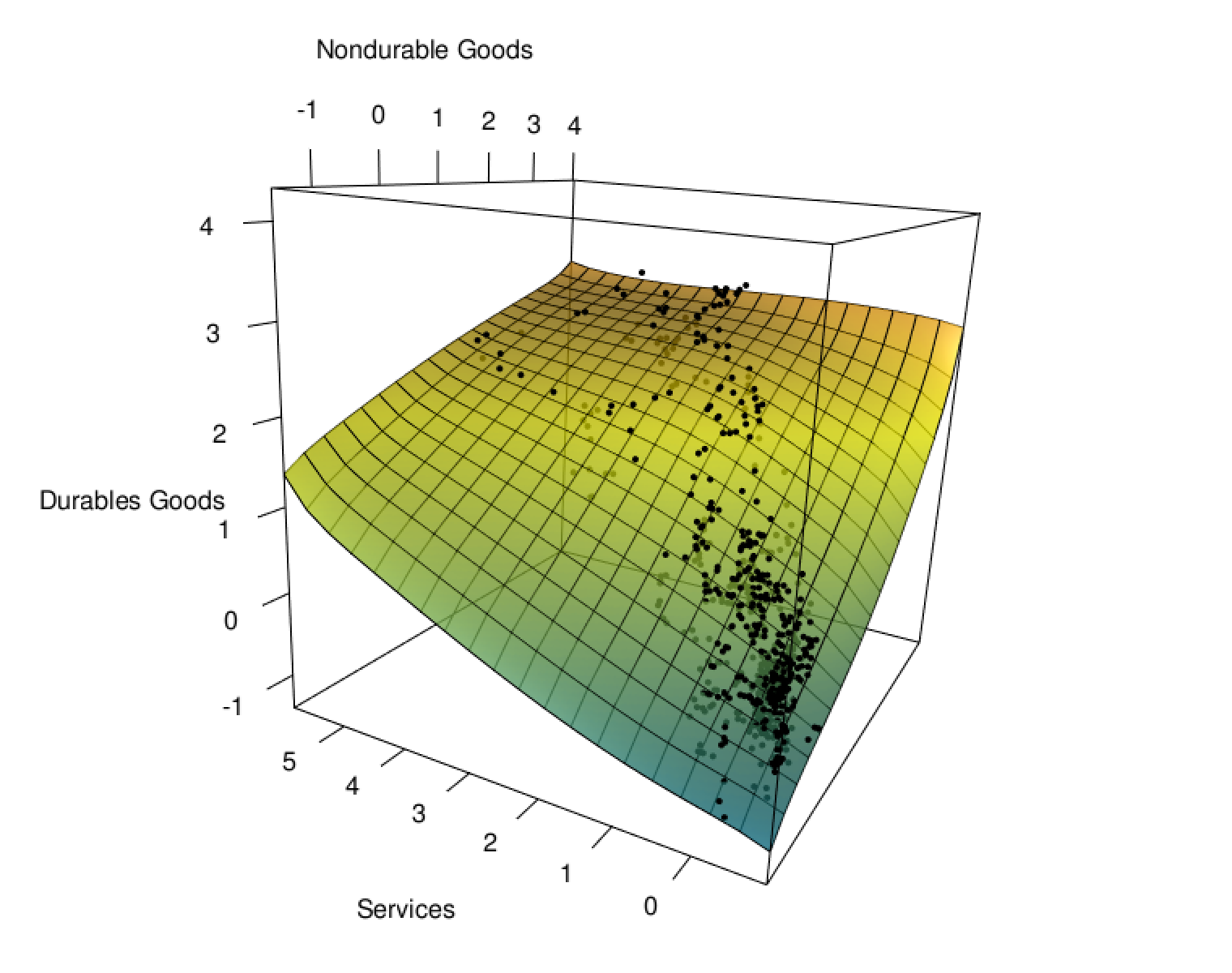}
        \caption*{Quantile surface, confidence level $p=0.5$}
    \end{minipage}
    \hfill
    \begin{minipage}{0.32\linewidth}
        \centering
        \includegraphics[width=\linewidth]{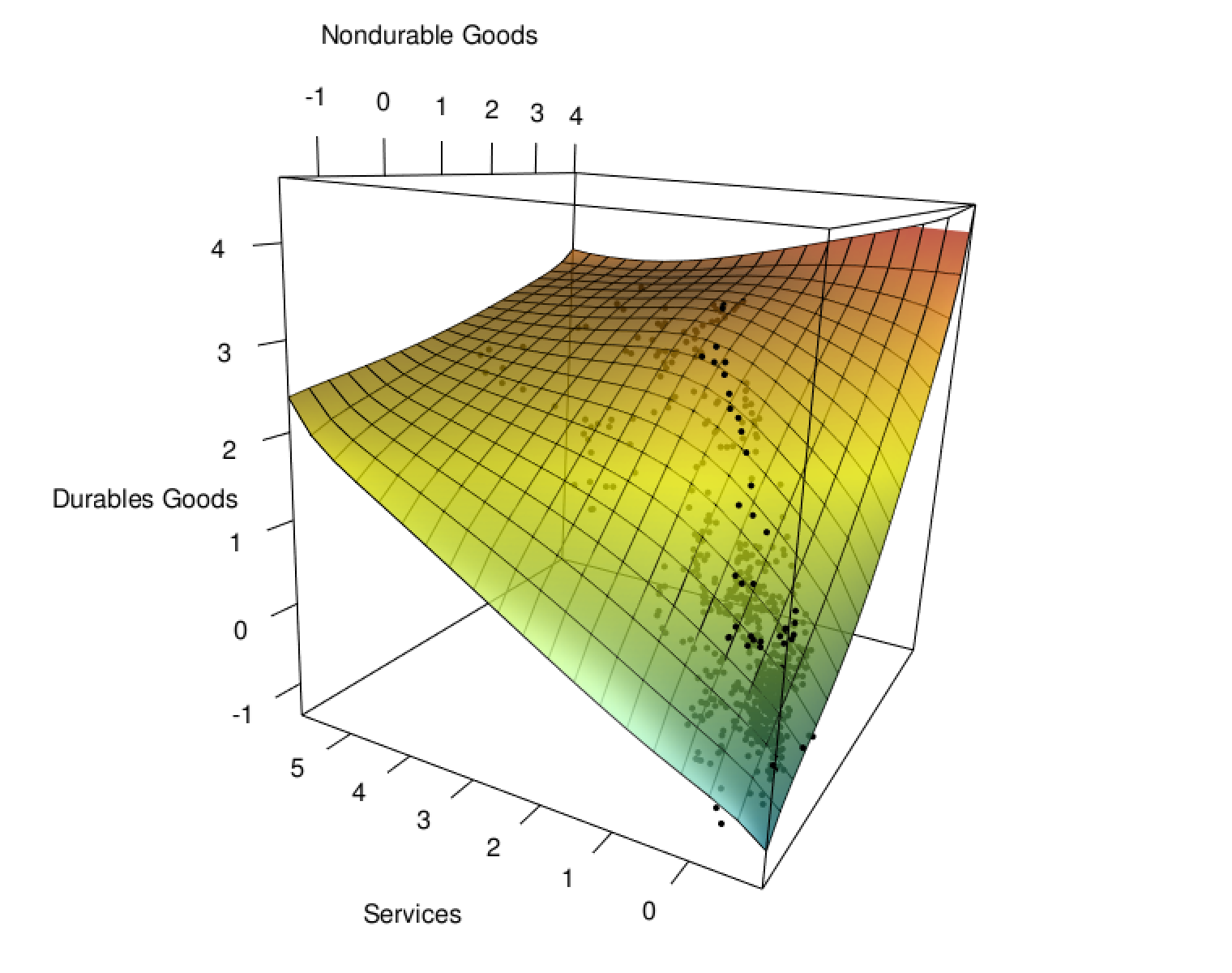}
        \caption*{Quantile surface, confidence level $p=0.95$}
    \end{minipage}
    \caption{Quantile surfaces of confidence levels $p=0.05,0.5,0.95$ in Experiment 2. The  dots are the data points. The quantile surface plots the conditional $p$-quantile of the response variable  as a function of the factors. 
    }\label{figure}
\end{figure*}

\begin{table*}
\caption{Average coverage rates of the prediction intervals  in 10-fold cross-validation}
\label{table_coverage}
\begin{tabular}{@{\extracolsep{5pt}} cccccccc} 
\\[-1.8ex]\hline 
\hline \\[-1.8ex] 
Model & 0.98 & 0.9 & 0.7 & 0.5 & 0.3 & 0.1 & Ave. Diff. \\ 
\hline \\[-1.8ex] 
Factor model of mixture quantiles & $0.980$ & $0.904$ & $0.706$ & $0.484$ & $0.279$ & $0.087$ & \bm{$0.007$} \\ 
Gaussian mixture regression & $0.986$ & $0.907$ & $0.690$ & $0.479$ & $0.277$ & $0.091$ & $0.013$ \\ 
Generalized additive model & $0.984$ & $0.920$ & $0.711$ & $0.492$ & $0.276$ & $0.090$ & $0.013$ \\ 
NGBoost & $0.970$ & $0.894$ & $0.696$ & $0.490$ & $0.288$ & $0.094$ & $0.008$\\
\hline \\[-1.8ex] 
\end{tabular} 
\caption*{Result of Experiment 1. The coverage rate is calculated by the percentage of the response variable that falls within the predicted conditional interval of two specified quantiles.  The average differences between the realized coverage rate and the target coverage rate are calculated. The target coverage rates are obtained by confidence intervals $(0.01,0.99)$, $(0.05,0.95)$, $(0.15,0.85)$,$(0.25,0.75)$,$(0.35,0.65)$,$(0.45,0.55)$.  }
\end{table*}

\section{DIMENSIONALITY REDUCTION BY REDUCED RANK TENSOR}\label{sec_dim_reduction}

This section introduces the basics of CP decomposition of tensors  and  an alternating algorithm for dimensionality reduction. 
Multivariate B-spline is obtained by tensor product of univariate B-splines. In high dimensions, the tensor product B-spline has a gigantic number of basis functions. 
  We propose an dimensionality reduction method by CP decomposition. The number of parameters of the reduced model grows linearly with respect to the dimension. It  can be viewed as creating new basis functions by optimal linear combination of original ones.   Our proposed alternating algorithm efficiently solves a smaller convex optimization problem in each iteration. 

\subsection{Tensor and CP Decomposition}

We adopt the following notations.
 $\bm{A}=$ parameter tensor.
$K=$ number of scalar factors.
$R=$ rank of parameter tensor.
$\otimes =$ tensor product of two vectors.
 $\bigotimes_{k =1}^\emph{K}=$ tensor product of $\emph{K}$ vectors.
 $<\cdot,\cdot>_F=$ Frobenius inner product of two tensors.

A tensor is a multidimensional array.  CP decomposition for a rank-$\emph{R}$ tensor $\bm{A}$ is defined as 
\begin{align}\label{cp_decomp}
\bm{A} = \sum_{r=1}^\emph{R}  
 \bm{u}_r^{(0)} \otimes \cdots \otimes \bm{u}_r^{(\emph{K})} \;,
\end{align}
where $\{\bm{u}_r^k\}_{r=1,k=0}^{\emph{R},\emph{K}}$ are vectors.  

The following equation is useful for subsequent interpretation of reduced rank method
\begin{align}\label{tensor_dot_product}
\bigotimes_{k =0}^\emph{K} \bm{u}_r^{(k)} 
\cdot 
\bigotimes_{k =0}^\emph{K} \bm{v}_r^{(k)}
=
\prod_{k=0}^\emph{K} 
 \bm{u}_r^{(\emph{k})} \cdot \bm{v}_r^{(\emph{k})}\;.
\end{align}

\subsection{Reduced Rank Method for Parameter Tensor}

We adopt the following notations. 
 $\emph{K}=$ number of scalar factors.
$\emph{L} =$ number of basis spline functions for each scalar factor. 
$\emph{J} = \emph{K} \times \emph{L}$.
 $\bm{A}=$ parameter tensor with the same elements as parameter vector $\bm{a}$.
 $\bm{B}_k(x) = (B_{k1}(x),\cdots,B_{k\emph{L}}(x))=$ vector of univariate spline basis of scalar factor $x_k$.
  $\bigotimes_{k =1}^\emph{K} \bm{B}_k(x_k) =$ tensor product of $\emph{K}$ vectors of univariate spline basis of scalar factors.
 $\bm{Q}(p) = (Q_0(p),\cdots,Q_\emph{I}(p))=$ vector of basis quantile functions.
$ \geq$  represents elementwise inequality when used on vector, matrix and tensor.

For B-spline, $\emph{L}=$ degree of univariate polynomial$+$number of knots$+1$.  Multivariate B-spline basis is obtained by tensor product of univariate B-spline basis.

With the above notations, we can write the model \eqref{model_2} in tensor format
\begin{align}
 G(p,\bm{x},\bm{a}) 
 =
 \left<
 \bm{A} , \,
 \bm{Q}(p) \otimes 
  \bigotimes_{k =1}^\emph{K} \bm{B}_k(x_k) 
  \right>_F
 \;.
\end{align}
Note that in our model formulation, $\bm{A}\geq 0$. Consider a nonnegative version of CP decomposition \eqref{cp_decomp} on the parameter tensor $\bm{A}$ where $\bm{u}_r^{(k)} \geq 0$.  With \eqref{cp_decomp}\eqref{tensor_dot_product}, we have
\begin{align*}\label{model_rank_decomp}
 G(p,\bm{x},\bm{a}) 
 =
  \sum_{r=1}^\emph{R}
\left(
\bm{u}_r^{(0)} \cdot \bm{Q} \right) 
\prod_{k=1}^K
\left(
  \bm{u}_r^{(k)} \cdot \widehat{\bm{B}}_K
 \right).
\end{align*}

Thus reduced rank method has a straightforward interpretation. The model is reduced to a linear combination of $\emph{R}$ functions from $(\emph{I}+1)(\emph{J}+1)$. In the extreme case where $\emph{R} = 1$, the model reduced to a heteroscedastic model where the conditional quantile function is $\bm{u}_r^{(0)} \cdot \bm{Q}$, whose  scale is controlled by a scalar function $\prod_{k=1}^\emph{K}
  \bm{u}_r^{(k)} \cdot \bm{B}_k$. It can be regarded as obtaining $\emph{R}$ new basis quantile functions $\{\bm{u}_r^{(0)} \cdot \bm{Q}\}_{r=1}^\emph{R}$, likewise for basis spline functions. The new basis, having a smaller number, can still have undesirable smoothness condition. We expect better performance when it is used along with penalties in P-spline.  The reduced basis functions are a linear combination of all original basis functions. This is in contrast to sparse optimization which leads to a subset of basis functions by forcing zeros among the parameters.  We find that the solution to Problem \ref{first_method} is often sparse with many small nonzero values. Thus the low rank decomposition is expected to produce good approximation,  although the decomposition \eqref{cp_decomp} is not always valid for any $\bm{A}$ for a low rank $\emph{R}$.

\subsection{Alternating Algorithm}

We propose an alternating algorithm to find $\bm{A}$. Let $\bm{U}_k = (\bm{u}_1^{(k)},\cdots,\bm{u}_\emph{R}^{(k)})$, $k=1,\cdots,\emph{K}$. 
Define $\widetilde{L}\left( \{\bm{U}_k\}_{k=0}^\emph{K};\bm{Y}^\emph{N},\bm{x}^\emph{N},p \right) =$  discrete residual random variable taking with equal probabilities the following values  for  $n=1,\ldots,\emph{N}$
\begin{equation*}
y_n -   
  \sum_{r=1}^\emph{R}
\left(
\bm{u}_r^{(0)} \cdot \bm{Q}(p) \right)
\prod_{k=1}^{K}
\left(
  \bm{u}_r^{(k)} \cdot \bm{B}_k(\bm{x}_n)
 \right)
\;.
\end{equation*}

Each step in the algorithm is a convex optimization problem. The objective function is nonincreasing in each step. The algorithm will find a local minimum at the end. Different initialization is needed for finding a smaller local minimum.

 \begin{algorithm}
  \caption{Alternating Algorithm for Estimating Reduced Rank Parameter Tensor}\label{algo}
  \begin{algorithmic}[1]
  \REQUIRE Data $\{(y_i,\bm{x}_i)\}_{i=1}^\emph{N}$;  confidence levels $\{p_m\}_{m=1}^\emph{M}$; model $G(p,\bm{\theta},{\bm{a}})$; error function $\bm{\mathcal{E}}_p$; threshold $\epsilon$
\STATE Initialize $\{\bm{U}_k\}_{k=0}^\emph{K}$
\REPEAT
  \FOR{$k=0$ to $\emph{K}$}
 \STATE Update $\bm{U}_k^{(s)}$ to $\bm{U}_k^{(s+1)}$ with fixed $\{\bm{U}_{k_t}^{(s)}\}_{k_t \neq k}$ 
\begin{equation*}\label{problem_statement_iterative_2}
\begin{aligned}
\bm{U}_k^{(s+1)} = &
\\
\argmin_{\bm{U}_k \geq 0} \sum_{m=1}^\emph{M}  w_m \bm{\mathcal{E}}_{p_m} & \left( \widetilde{L}( \{\bm{U}_k^{(s)}\}_{k=0}^\emph{K} ; \bm{Y}^\emph{N},\bm{x}^\emph{N},p_m  ) \right) 
\end{aligned}
\end{equation*}
 \ENDFOR
 \STATE $s\,+=1$
 \UNTIL{
 $
 ||\bm{A}^{(s+1)} - \bm{A}^{(s)}||_F \leq \epsilon
 $
 }
\ENSURE Parameter tensor $\bm{A}$
  \end{algorithmic}
  \end{algorithm}

\section{NUMERICAL EXPERIMENTS}\label{sec_numerical}    
\paragraph{Experiment 1} 
This experiment  compares the performance of our model\footnote{The R implementation is available at \url{https://github.com/ch-pg/distributional_regression}.} with three benchmarks. 
The dataset is Combined Cycle Power Plant from UCI Repository ($4$ features, $1$ target and $9568$ data points).  We fit our model and three benchmarks in $10$-fold cross-validation and compare  CRPS and coverage rate. The coverage rate is calculated by the percentage of the response variable that falls within the predicted conditional interval of two specified quantiles. The detailed setup of all experiments is in  Appendix \ref{sec_experiment_append}.

The results are summarized in Table \ref{table_coverage} and Table \ref{table_crps}. Our approach demonstrates an improved average out-of-sample coverage rate. The average out-of-sample CRPS for our method is lower than that of GMR and GAMLSS but higher than NGBoost. Our approach has the shortest fitting time.

\begin{table}
\centering
\caption{CRPS  and time cost}
\begin{tabular}{ccccc}
\label{table_crps}
\\[-1.8ex] 
\hline \\[-1.8ex] 
Model & FMMQ & GMR & GAMLSS & NGBoost \\ \hline
CRPS & $0.080$ & $0.088$ &  $0.081$ & \bm{$0.075$} \\
Std CRPS & \bm{$0.002$} & $0.003$ & $0.003$ & \bm{$0.002$} \\
Time(s) & \bm{$11.1$} & $109.0$ & $29.0$ & $12.0$ \\ \hline

\end{tabular}
\caption*{FMMQ = Factor model of mixture quantiles, GMR = Gaussian mixture regression, GAMLSS = generalized additive models for location, scale and shape.
std CRPS = standard deviation of CRPS in 10-fold cross-validation. Time(s) = average time cost in seconds.
}
\end{table}

\paragraph{Experiment 2}
Figure \ref{figure} visualizes the quantile surfaces of confidence levels $q=0.05,0.5,0.95$.  
The dataset is monthly CPI of three categories \citep{mccracken2016data,Fredcpi}.

\paragraph{Experiment 3} 

The experiment tests the convergence of the alternating algorithm and loss of fit of the dimensionality reduction method for models with rank $1$, $3$, and $10$. 
We calculate the log of reduction in the objective function during training, i.e., initial training loss  minus training loss at current step. 

\begin{figure}[]
\centering
\includegraphics[width = 1\linewidth]{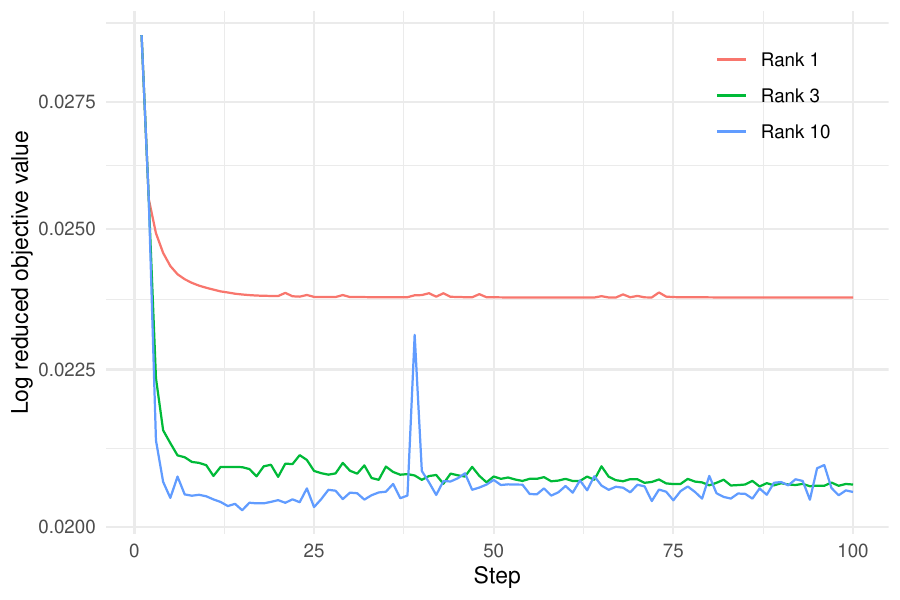}
\caption{Experiment 3: Semilog plot of training loss using Algorithm \ref{algo} on models with rank $1$, $3$ and $10$.}
\label{fig_obj}
\end{figure}

The log reduced objective value at each step is displayed in Figure \ref{fig_obj}. We observe that in all cases, the value converges in less than 20 iterations, demonstrating the effectiveness of the proposed algorithm. The optimal objective values of models with rank 3 and 10 are very close, while the model with rank 1 has a larger objective value as expected. The nonmonotonicity of the curves is caused by numerical errors.

\section{CONCLUSION}

The proposed  nonparametric distributional regression has a strong approximation property, can be efficiently estimated by convex optimization, and allows for  dimensionality reduction by tensor decomposition. We analysis the model specifications and equivalent formulations, and propose an alternating algorithm for computation.

\nocite{cvar1,cvar2,ghahramani1996mixture}
\bibliographystyle{chicago}
\bibliography{FM_reference.bib}{}

\onecolumn
\appendix
\numberwithin{equation}{section}
\aistatstitle{Supplementary Materials}
\section{Detailed Literature Review}\label{sec_review_append}


\textit{Mixture quantiles model}  uses a combination of some basis functions  functions to model the quantile function. The idea is studied in \cite{tukey1962future, gilchrist2000statistical}.  Various basis functions have been used, such as orthogonal polynomials \citep{Sillitto1969}, a mixture of normal and Cauchy distributions with linear and quadratic terms \citep{Karvanen2006mixtures}, and a modified logistic distribution \citep{Keelin2016metalog}. \cite{vincent1912vincent,
ratcliff1979group} have a different motivation to obtain population quantile function by averaging  individual quantile functions. 
However, these methods lack factor dependence and some do not guarantee that the resulting quantile function is nondecreasing. 
 Our model  uses mixture quantiles as the conditional quantile function (Section \ref{sec_model}). Note that mixture quantiles and mixture densities are two different families of distributions.

 \textit{Mixture regression} shares a similar idea to our approach, where the conditional distribution is modeled as a linear combination of basis functions. 
   The conditional density function of the widely applied Gaussian mixture regression is modeled  by a mixture (weighed sum) of Gaussian densities, where the weight, mean and variance  of each component can be modeled as functions of  factors. 
    Gaussian mixture regression has been studied under different names, such as switching regression model, mixture of experts  \citep{Quandt1972switching,
desarbo1988regression,
VILLANI2009155mixture,
Yuksel2012reviewmixture}. 
Similarly, in our model formulation,  the weights of quantile functions are functions of the factors (Section \ref{sec_model}). Mixture regression is generally computationally expensive to estimate, while our parameter estimation can be solved by convex optimization.  The weight function in mixture regression must sum up to one, which is typically achieved by the softmax transformation, while our model only requires nonnegative weight functions.

\textit{Quantile regression} 
estimates the conditional quantile of a given confidence level given  some factors. 
The application of spline functions to conduct nonparametric quantile regression has been presented in  \cite{koenker1994spline,He1998bispline,Koenker2011additive}. However, conducting multiple quantile regressions separately can result in quantile crossing, particularly in nonlinear models, which impedes the interpretation of the results. To mitigate this problem, various methods have been proposed, such as imposing extra constraints \citep{bondell2010noncrossing} or rearrangement \citep{Chernozhukov2010rearrange}. The constraints are often imposed at observed data points or at some confidence levels.  
We present a novel approach to quantile regression that guarantees noncrossing quantiles without any post-processing. Our method estimates the entire conditional distribution of the response variable, not just the quantiles at several confidence levels. The proper choice of basis functions ensures the validity of the noncrossing property at all confidence levels (Section \ref{sec_model}).  The objective function in our approach is a linear combination of pinball losses at different confidence levels (Section \ref{sec_calibration}), which has been adopted in various parameter estimation methods \citep{KOENKER1984wqr,
Zou2008additive,
Sottile2022wqr}.

\textit{Quantile regression process} refers to the regression coefficient as a function of the confidence level of quantile regression.  \cite{Angrist2006qrpasymptotics} shows that the rescaled quantile regression process converges to a zero-mean Gaussian process.   
A number of independent but closely related studies have explored  the use of various basis functions for modeling the quantile regression process, such as Bernstein basis polynomials  \citep{Reich2011Bqrprocess}, P-spline basis \citep{LIAN2015qrprocess},  common  parameterized functions \citep{Frumento2016WQR}, and   monotone B-spline \citep{Yuan2017Splineqrprocess}. 
By rearranging the model formulation  in Section \ref{sec_model} and viewing each polynomial term as a factor, 
it can be shown that our model coincides with the quantile regression process models.

There are several notable differences between our model and existing models on quantile regression process. First, the parameter estimation in \cite{Frumento2016WQR} and \cite{Reich2011Bqrprocess} involves numerical integration and Bayesian method, respectively, while our approach uses convex optimization, which is more efficient. Second, 
it it can be challenging to ensure noncrossing quantiles when directly modeling the linear quantile regression process. \cite{Reich2011Bqrprocess} introduces  prior latent unconstrained variables; 
\cite{Frumento2016WQR} checks the nonnegativity of derivative; 
 \cite{Yuan2017Splineqrprocess} uses linear constraints when the feasible set is a bounded convex polytope;  \cite{LIAN2015qrprocess} does not guarantee noncrossing quantiles.  In contrast, our model guarantees noncrossing conditional quantiles at any two distinct confidence levels and any given factor value.  Third,  using a spline to model quantile regression process results in bounded conditional quantile function, which may lead to severe underestimation of tail risk. Thus it is important to supplement splines with quantile functions of common distributions. 
 Furthermore, all aforementioned quantile regression process models study linear models and can be regarded as special cases of our approach. To ensure that a linear quantile model does not have quantile crossing given any factor values, the coefficients of each factor must be equal for all confidence levels. This condition leads to a homoscedastic model.

  \textit{Continuous Ranked Probablity Score} (CRPS, \cite{Matheson1976score}) is a proper scoring rule  that is frequently used to measure the quality of a distributional prediction when the response variable is  a scalar. 
 \cite{Laio2006VerificationTF} proposes an equivalent definition that is useful to formulate convex optimization.  \cite{Hothorn2014Transformation,
Gouttes2021NN,BERRISCH2021CRPS} use CRPS as the objective function of optimization in learning tasks. 
\cite{Zhang2022CDFmixture} proposes a related model where the CDF is a linear combination of basis CDFs.  Although not mentioned in the paper, their estimation procedure minimizes CRPS.

\textit{Quantile model aggregation} has a similar model formulation, but aims to aggregate multiple conditional quantile models to improve overall performance. The aggregation can be performed across different factor values and different quantile levels.  \cite{BERRISCH2021CRPS} uses B-splines as weight function across confidence levels, while \cite{
Fakoor2021Aggregation} uses neural networks as weight function across different factor values and different confidence levels. Viewing each model as a basis function, our approach can be immediately adapted  for quantile model aggregation across different factor values. 
 On the other hand, \cite{Fakoor2021Aggregation} considers a regression model called the deep quantile regression  by using constant functions as individual models. The conditional quantile function is a neural network, which requires extra efforts to ensure monotonicity.

\textit{Conditional transformation model} \citep{Hothorn2014Transformation}   finds the  optimal conditional spline transformation of the conditional distribution, such that the quantile function of the transformed conditional distribution can be modeled by a predetermined quantile function. 
In contrast, our approach can be viewed as modeling the inverse of the conditional transformation. 
 Our model is also related to the \textit{generalized additive models for location, scale and shape} (GAMLSS,  \cite{Rigby2005gamlss}). When there is only one basis quantile function, e.g., quantile function of normal distribution, the mean and variance are modeled as splines of the factor, which is equivalent to generalized additive models for location and scale. 
 \textit{The linear and polynomial model} in \cite{chernozhukov2001conditional} can be regarded as a special case of our model.  
 The extension of our approach to neural networks is related to
\textit{Spline quantile function RNN} \citep{Gasthaus2019quantile}, which models the parameter of piecewise linear splines with neural networks and estimate the parameters by CRPS minimization. Similarly, the model ensemble method in \cite{Bremnes2020polyNN} uses Bernstein polynomials.

Multivariate B-spline is obtained by tensor product of univariate B-splines. The CP decomposition for spline tensors has been applied in Computer Aided Design (CAD) to reduce model complexity \citep{Pan2016cad}. For a comprehensive review of tensor decomposition, the readers are referred to \cite{Kolda2009tensor_intro, Rabanser2017tensor_intro}. This study applies the technique to statistical estimation. In high dimensions, the tensor product B-spline has a gigantic number of basis functions. The dimensionality reduction method by CP decomposition can be viewed as creating new basis functions by optimal linear combination of original ones. 
Our approach is different from existing work on dimensionality reduction for quantile regression. 
  \cite{Lian2019reducedrank} proposes reduced matrix rank regression for  homoscedastic multiple quantile modeling.  \cite{Chen2021QFM} proposes a different approach to dimensionality reduction for quantile regression determining different principle components for different confidence levels.  Our approach also results in different factors for different confidence levels (Section \ref{sec_dim_reduction}).

Distributional estimation has been extensively studied. 
\textit{Conformal prediction} \citep{vovk2005algorithmic} post-processes a point forecast to obtain prediction interval with guarantees on coverage. \textit{Conformalized quantile regression}  \citep{romano2019conformalized} conducts quantile regression, then conformalizes the quantile estimate. Our method, when focused on a specific confidence level, can be conformalized using \cite{romano2019conformalized}. 

Tree-based methods  and neural networks are very effective on tabular data, see for example \cite{meinshausen2006quantile,
mccarter2024unmasking,
jolicoeur2024generating,
ram2011density,
 beltran2024treeffuser}. However, they often do not provide a closed-form expression for the conditional quantile function. \textit{NGBoost} \citep{duan2020ngboost} proposes a modular boosting approach consisting of  base learner, parametric distribution, and score function. Our model formulation is equivalent to NGBoost model formulation when  the base learner is the basis weight functions $\{{a}_j B_j(\bm{x_n})\}_{j=1}^J$ with unknown parameter ${a}_j$; the parametric distribution has quantile function $\sum_{i=1}^I a_i Q_i(p)$; the scoring rule is CRPS. Similarly, the distribution parameters are obtained by an additive combination of base learner outputs, that is,  $a_i = \sum_{j=1}^J a_{ij} B(\bm{x_i})$.

\section{Proof of Theorem \ref{theorem_approx}}\label{sec_proof}
\begin{proof} 
We want to prove that for any  $ G(p,\bm{x},\bm{\theta}) \in G_{[0,1]}$ and any $\epsilon>0$, there exists  $F(p,\bm{x}) \in \cup_i \mathcal{P}(Q_i(p)  \otimes B_j(\bm{x}), T_i)$ such that 
$$|F(p,\bm{x}) - G(p,\bm{x},\bm{\theta})| <\epsilon$$ for $\bm{x} \in \mathcal{X}$. 
Note that $\forall G(p,\bm{x},\bm{\theta}) \in G_{[0,1]\times \mathcal{X}}$, we have $\frac{\partial}{\partial p}G(p,\bm{x},\bm{\theta}) \geq 0$, since quantile function is nondecreasing in $p$.  
By Theorem 4.1 of \cite{Papp2011polynomial}, for any $\epsilon>0$, there exists a tensor-product B-spline $f(p,\bm{x})$ with nonnegative coefficients  such that $$|\frac{\partial}{\partial p}G(p,\bm{x},\bm{\theta}) - f(p,\bm{x})| < \epsilon .$$
  Integrating $f(p,\bm{x})$ with respect to $p$, we obtain
$$\int f(p,\bm{x}) \mathrm{d}p = 
\widetilde{F}(p,\bm{x}) + \widetilde{t}(\bm{x}) ,$$ where $t(\bm{x})$ is a differentiable function. 
Since I-spline is defined by the integral of B-spline, we have $\widetilde{F}(p,\bm{x}) \in \cup_i \mathcal{P}(Q_i(p)  \otimes B_j(\bm{x}), T_i)$.  
For any $ \frac{1}{2}\epsilon > 0$, there exists a B-spline $t(\bm{x})$ such that $$|\widetilde{F}(0.5,\bm{x}) + t(\bm{x})  - G(0.5,\bm{x},\bm{\theta})| < \frac{1}{2}\epsilon .$$ 
Let $\widetilde{t}(\bm{x}) = t(\bm{x})$.  
We have $\widetilde{F}(p,\bm{x}) + \widetilde{t}(\bm{x}) \in \cup_i \mathcal{P}(Q_i(p)  \otimes B_j(\bm{x}), T_i)$. We also have 
\begin{align*}
& |\widetilde{F}(p,\bm{x}) + \widetilde{t}(\bm{x}) - G(p,\bm{x},\bm{\theta}) - (\widetilde{F}(0.5,\bm{x}) + \widetilde{t}(\bm{x}) - G(0.5,\bm{x},\bm{\theta}))|  \\
< &
\max\left\{
\int_0^{\frac{1}{2}} |\frac{\partial}{\partial p}G(p,\bm{x},\bm{\theta}) - f(p,\bm{x})| \mathrm{d}p ,
\int_{\frac{1}{2}}^1 |\frac{\partial}{\partial p}G(p,\bm{x},\bm{\theta}) - f(p,\bm{x})| \mathrm{d}p
 \right\}\\
< & \frac{1}{2} \epsilon
\end{align*} 
Hence for any $\bm{x} \in \mathcal{X}$, we have $$|\widetilde{F}(p,\bm{x}) + \widetilde{t}(\bm{x}) - G(p,\bm{x},\bm{\theta})|< \frac{1}{2} \epsilon + \frac{1}{2} \epsilon = \epsilon .$$ 
\end{proof}

\section{Asymptotic Property}\label{sec_asym}

This section contains the asymptotic properties of the estimator. We adopt the following notations.
$\widehat{\mathcal{E}}_\emph{N}(\bm{a})  =  \sum_{m=1}^\emph{M}  w_m \bm{\mathcal{E}}_{p_m} \left( L(\bm{a};\bm{Y}^\emph{N},\bm{x}^\emph{N},p_m) \right)$.
 $\widehat{\bm{a}}_\emph{N} = \argmin_{\bm{a} \in \mathcal{A}} \widehat{\mathcal{E}}_\emph{N}(\bm{a})$.
 ${\bm{a}}_0=$ true parameter.

Theorem \ref{theorem_consistency}, \ref{theorem_normalilty} in the following from \cite{Frumento2016WQR} are applications of \cite{Newey2}.

\begin{theorem}\label{theorem_consistency}
Assume that $\mathcal{A}$ is a compact set. If there is a function $\widehat{\mathcal{E}}_0(\bm{a})$  such that (i) $\widehat{\mathcal{E}}_N(\bm{a})$ converges uniformly in probability to $\widehat{\mathcal{E}}_0(\bm{a})$; (ii) $\widehat{\mathcal{E}}_0(\bm{a})$ is uniquely minimized by $\widehat{\bm{a}}$; (iii) $\widehat{\mathcal{E}}_0(\bm{a})$ is continuous at $\widehat{\bm{a}}$.  Then $\widehat{\bm{a}}_\emph{N} \overset{d}{\rightarrow} {\bm{a}}_0$.
\end{theorem}

\begin{theorem}\label{theorem_normalilty}
Suppose that the conditions Theorem \ref{theorem_consistency} are satisfied, and (i) $\bm{a}_0$ is an interior point of $\mathcal{A}$; (ii) $\widehat{\mathcal{E}}_\emph{N}(\bm{a})$ is twice continuously differentiable in a neighborhood  of $\bm{a}_0$; (iii) $\sqrt{n} \nabla_{\bm{a}} \widehat{\mathcal{E}}_\emph{N} (\bm{a}_0) \overset{d}{\rightarrow} \mathcal{N}(\bm{0},\Omega)$; (iv) there is $H(\bm{a})$ that is continuous at $\bm{a}_0$ and $\sup_{\bm{\theta}_0 \in \mathcal{A}} ||\nabla_{\bm{a}\bm{a}} \widehat{\mathcal{E}}_\emph{N} (\bm{a}_0) - H(\bm{a}) || \overset{p}{\rightarrow} \bm{0}$; (v) $H = H(\bm{a}_0)$ is nonsingular. Then
 \begin{align}
\sqrt{N}(\widehat{\bm{a}}_\emph{N} -\bm{a}_0) \overset{d}{\rightarrow} \mathcal{N}(\bm{0},H^{-1} \Omega H^{-1} ) \;.
\end{align}
\end{theorem}

\section{Details of Numerical Experiments}\label{sec_experiment_append}

Experiments were conducted on a 13-inch MacBook Pro (2020) equipped with a 2 GHz Quad-Core Intel Core i5 processor, 16 GB of 3733 MHz LPDDR4X memory, and integrated Intel Iris Plus Graphics with 1536 MB of VRAM, running macOS Sequoia 15.1.1.

\subsection{Experiment 1}

\paragraph{Data}

The dataset is Combined Cycle Power Plant from UCI Repository. The data are standardized to have zero median and unit interquartile range. The data is randomly shuffled before experiments.

\paragraph{Model}
Based on preliminary examination of the dataset, we choose the following model for efficient computation. 
\begin{itemize}
\item $Q_\mathtt{N}(p) = \sqrt{2}{\erf}^{-1}(2p-1) = $ quantile function of standard normal distribution
\end{itemize}
Consider the Two-Factor Model of Mixture Quantiles 
\begin{equation}
\begin{aligned}\label{model_experiment_1}
G(p,\bm{x},\bm{a})
=
 &f_0(\bm{x}) + 
  Q_\mathtt{N}(p) f_1(\bm{x}) \;, 
\end{aligned}
\end{equation}
 For the B-splines $f_0, f_1$, we use basis of degree $2$ with $3$ equidistant knots. Hyperparameter of the model includes the number of basis quantile functions, the degree of basis weight functions, the position of knots, etc. Hyperparameter tuning is not pursued in this study.

\paragraph{Parameter Estimation} The confidence levels used in the optimization are $0.01, 0.33, 0.66, 0.99$. The weights for the confidence levels are $0.5, 0.2, 0.2, 0.5$ and normalized such that they sum up to $1$.  
We add a penalty term on the squared difference between adjacent coefficients \citep{Eilers1996Pspline,
Eilers2003multiPspline} for the spline function $f_1$ to reduce overfit. The penalty coefficients are fixed at $0.1$, which is not optimized by cross-validation.  
 The optimization for parameter estimation is conducted with R package Portfolio Safeguard \citep{zabarankin2016book}.

\paragraph{Benchmarks} We choose  the Gaussian mixture regression,  the generalized additive model (GAM) and NGBoost  as the benchmarks. For the Gaussian mixture regression, we use $3$ mixture components. The weights of the components are constants in the model. For the generalized additive model, the mean, standard deviation of the normal distribution  are functions of the factors. The additive formula of the functions are  sum of univariate cubic splines of each factor. Adding  additional interaction terms often leads to convergence issue. The univariate cubic splines are  penalized on the second derivative of the functions. For the three modules in NGBoost, we use trees with depth $3$ as the base learner, normal distribution as the parametric function, and CRPS as the score function.   The benchmark models are implemented by R packages Mixtools \citep{Benaglia2009mixtools}, GAMLSS \citep{Rigby2005gamlss} and Python library NGBoost, respectively.

\paragraph{Performance Measure} We conduct 10-fold cross-validation and use  the out-of-sample CRPS and coverage rate as performance measures. The CRPS is calculated by sum of CRPS of distributional prediction on response variable. The coverage rate is calculated by the percentage of the response variable that falls within the predicted conditional interval of two specified quantiles.  A small CRPS and a coverage rate close to the target indicate a high-quality distributional prediction.

\paragraph{Results} Results are presented in the main part of the paper.

\subsection{Experiment 2}




 The experiment focuses on the Factor Model of Mixture Quantiles (Section \ref{sec_model}) and its parameter estimation by convex optimization (Section \ref{sec_calibration}). The model is used to estimate the conditional distribution of Durable Goods as a function of Nondurable Goods and Services. 
 
 
\paragraph{Data}
The response variable is the percentage change from previous year of the Consumer Price Index for All Urban Consumers: Durables in US City Average, while the factors are the percentage change from previous year of the Consumer Price Index for All Urban Consumers: Nondurables in US City Average and the percentage change from previous year of the Consumer Price Index for All Urban Consumers: Services in US City Average. For convenience, we denote them by Durable Goods, Nondurable Goods and Services in subsequent sections. The monthly data, covering the period from January 1957 to January 2023, is described in \cite{mccracken2016data} and downloaded from \cite{Fredcpi}. The data are standardized to have zero median and unit interquartile range. The data is randomly shuffled before experiments.

\paragraph{Model}
\begin{itemize}
\item $Q_\mathtt{N}(p) = \sqrt{2}{\erf}^{-1}(2p-1) = $ quantile function of standard normal distribution
\item $Q_{\mathtt{E}_1}(p) = - \ln(4-4p)=$ quantile function of right-side transformed exponential distribution that is nonzero in $(0.75,1)$
\item $Q_{\mathtt{E}_1}(p) =  \ln(4p)=$ quantile function of left-side transformed exponential distribution that is nonzero in $(0,0.25)$
\end{itemize}
Consider the Two-Factor Model of Mixture Quantiles 
\begin{equation}
\begin{aligned}\label{model_experiment}
G(p,\bm{x},\bm{a})
=
 &f_0(x_1,x_2) + 
  Q_\mathtt{N}(p) f_1(x_1,x_2) 
  \\
  &+ 
  Q_{\mathtt{E}_1}(p) f_2(x_1,x_2) +
  Q_{\mathtt{E}_2}(p) f_3(x_1,x_2) \;, 
\end{aligned}
\end{equation}
 For each bivariate B-spline $\{f_i\}_{i=0,1,2,3}$, we use basis of degree three with $10$ equidistant knots. 
The interpretation is that $f_1(x_1,x_2)$ determines the body of the distribution, while $f_2(x_1,x_2)$ and $f_3(x_1,x_2)$ determine the left and right tails, respectively.

\paragraph{Parameter Estimation} The confidence levels used in the optimization are $0.01,0.05,0.15,0.25,\cdots,0.95,0.99$. The weights for the confidence levels are $20,10,1,,:\cdots,1,10,20$ and normalized such that they sum up to $1$.  
We add a penalty term on the squared difference between adjacent coefficients \citep{Eilers1996Pspline,
Eilers2003multiPspline} for each spline function $\{f_i\}_{i=0,1,2,3}$ to reduce overfit. The penalty coefficients are fixed at $0.01$ and not optimized. 
 The optimization for parameter estimation is conducted with R package Portfolio Safeguard \citep{zabarankin2016book}.

\paragraph{Benchmarks} We choose  the Gaussian mixture regression and  the generalized additive model (GAM) as the benchmarks. For the Gaussian mixture regression, we use BIC as the model selection criteria, which is one of the built-in methods in the package Mixtools. The weights of the components are constants in the model. For the generalized additive model, the mean, standard deviation, shape and skewness   of the Skew $t$ type 2  distribution \citep{Azzalini2003student} are functions of the factors. The additive formula of the functions are  sum of univariate cubic splines of each factor  with an additional linear interaction term. The univariate cubic splines are  penalized on the second derivative of the functions.   The benchmark models are implemented by R packages Mixtools \citep{Benaglia2009mixtools} and GAMLSS \citep{Rigby2005gamlss}, respectively.

\paragraph{Performance Measure} We conduct 10-fold cross-validation and use  the out-of-sample CRPS and coverage rate as performance measures. The CRPS is calculated by sum of CRPS of distributional prediction on response variable. The coverage rate is calculated by the percentage of the response variable that falls within the predicted conditional interval of two specified quantiles.  A small CRPS and a coverage rate close to the target indicate a high-quality distributional prediction.

\paragraph{Results} Results are presented in Table \ref{appendix_table_coverage} and Table \ref{appendix_table_crps}. In general, our approach has comparable performance with the benchmarks. Our approach exhibits improved average out-of-sample coverage rate. The average out-of-sample CRPS of our approach is  lower than the Gaussian mixture regression but higher than the generalized additive model. However, the generalized additive model often yield poor extrapolation results when the test data lies outside the domain of training data, resulting in nonexistent second moments and therefore nonexistent CRPS values.

%

\begin{table*}
\caption{Average coverage rates of the prediction intervals  in 10-fold cross-validation for three  models}
\label{appendix_table_coverage}
\begin{tabular}{@{\extracolsep{5pt}} cccccccc} 
\\[-1.8ex] 
\hline \\[-1.8ex] 
Model  & 0.98 & 0.9 & 0.7 & 0.5 & 0.3 & 0.1 & Ave. Diff. \\ 
\hline \\[-1.8ex] 
Factor model of mixture quantiles & $0.977$ & $0.895$ & $0.741$ & $0.508$ & $0.297$ & $0.089$ & $\bf{0.005}$ \\ 
Gaussian mixture regression & $0.969$ & $0.907$ & $0.789$ & $0.603$ & $0.372$ & $0.113$ & $0.046$ \\ 
Generalized additive model & $0.975$ & $0.903$ & $0.728$ & $0.527$ & $0.315$ & $0.112$ & $0.013$ \\ 
\hline \\[-1.8ex] 
\end{tabular}
\caption*{Result of Experiment 1. The coverage rate is calculated by the percentage of the response variable that falls within the predicted conditional interval of two specified quantiles.  The average differences between the realized coverage rate and the target coverage rate are calculated. The target coverage rates are obtained by confidence intervals $(0.01,0.99)$, $(0.05,0.95)$, $(0.15,0.85)$,$(0.25,0.75)$,$(0.35,0.65)$,$(0.45,0.55)$.  }
\end{table*}

\begin{table*}[!htbp]
\centering
\caption{Average out-of-sample Continuous Ranked Probability Score (CRPS) for three  models: Factor model of mixtures, Gaussian mixture regression, and generalized additive model, obtained by  10-fold cross-validation procedure. The mean and standard deviation of the CRPS values are reported for each model. The asterisk $*$ denotes instances in which a model produces predictions of nonexistent second moment and therefore nonexistent CRPS values for some data points.  The CRPS is calculated only for predictions that have finite CRPS values. }\label{appendix_table_crps}
\begin{minipage}{1\linewidth}
\centering
\begin{tabular}{ccccccc}
\\[-1.8ex]
\hline \\[-1.8ex] 
Model & 1 & 2 & 3 & 4 & 5 & 6 \\ \hline
Factor model of mixture quantiles & 0.199 & 0.241 & 0.236 & 0.220 & 0.204 & 0.219 \\
Gaussian mixture regression & 0.209 & 0.262 & 0.247 & 0.238 & 0.215 & 0.241 \\
Generalized additive model & 0.186 & 0.213 & 0.210 & 0.204 & 0.193* & 0.208* \\ \hline
\end{tabular}
\end{minipage}%
\\
\begin{minipage}{1\linewidth}
\centering
\begin{tabular}{ccccccc}
\\[-1.8ex]
\hline \\[-1.8ex] 
Model & 7 & 8 & 9 & 10 & Mean & Std \\ \hline
Factor model of mixture quantiles & 0.197 & 0.255 & 0.193 & 0.186 & 0.215 & 0.023 \\
Gaussian mixture regression & 0.205 & 0.268 & 0.214 & 0.188 & 0.229 & 0.027 \\
Generalized additive model & 0.187* & 0.226 & 0.176 & 0.253 & 0.206 & 0.023 \\ \hline
\end{tabular}
\end{minipage}
\end{table*}

We can obtain the $p$-quantile surface of the response variable by varying the values of $(x_1,x_2)$ with a fixed $p$. We use the model fit in the $10$th fold of the cross-validation as an example. Figure \ref{figure} shows three selected quantile surfaces for visualization, demonstrating the nonlinear relations incorporated in the model. The surfaces do not to cross each other.

\subsection{Experiment 3}

\paragraph{Data} We randomly select two thirds of the data in Experiment 1.

 \paragraph{Model} We use $I$-spline basis of degree three with five knots and exponential distribution as basis quantile functions. for splines $\bm{B}_k(x_k)$ in weight function, we still use B-spline of degree three with six knots.

 \paragraph{Parameter estimation} 
 The initial parameter for three models are the same. 
 No penalty is applied to the smoothness of splines in this experiment, since the aim is to compare the convergence speed and loss of precision using varying ranks. 
 The nonnegativity constraint is applied in each step of the algorithm. However, there should be no sign constraint on the coefficients $\bm{a}_0$ of the spline $f_0(\bm{x},\bm{a}_0)$ that determines the conditional location, as shown in \eqref{model_expand}. To address this, we add a positive constant to all observations before applying the alternating algorithm. This ensures that the model $G(p,\bm{x},\bm{a})$ is positive, and  $\bm{a}_0$ satisfy the nonnegativity constraint. 
 We stop the iteration at step $100$. 
Other stopping rules can be used as well, such as the number of iterations, the difference between consecutively updated objective values. 

Note that the low-rank representation alone does not guarantee improvements in out-of-sample test, since the smoothness condition is unconstrained and may lead to overfit. The combination of the low-rank representation with P-spline is not pursued in this experiment. Due to this reason, we report the out-of-sample error during training only in the appendix for reader's reference. The validation set is the unselected one third of the data from Experiment 1.

\begin{figure}[h]
\centering
\includegraphics[width = 0.5\linewidth]{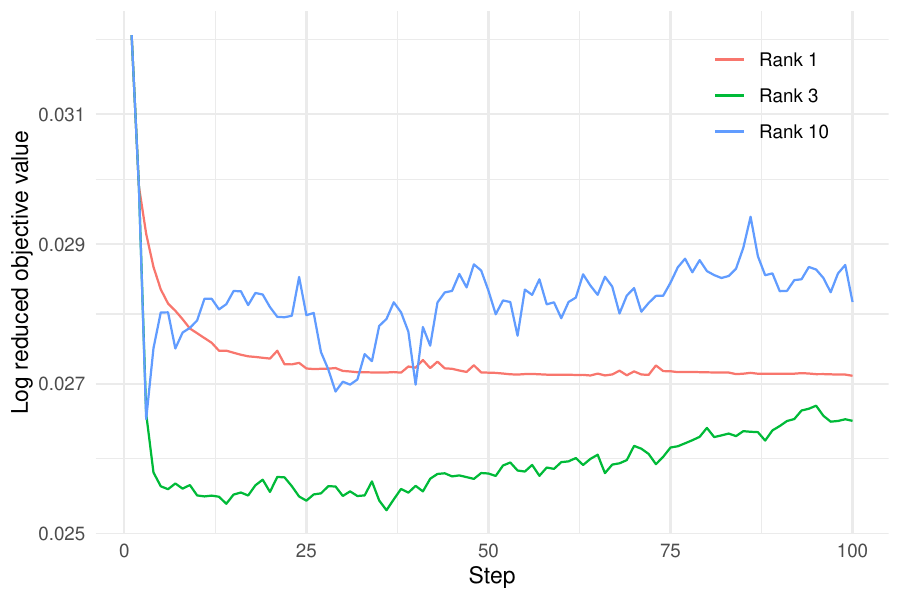}
\caption{Experiment 2: Semilog plot of validation loss using alternating algorithm on models with rank $1$, $3$ and $10$.}
\label{fig_quantile_tensor_oos}
\end{figure}

We observe in Figure \ref{fig_quantile_tensor_oos} that the out-of-sample errors decrease fairly fast in the first few iterations. Then, models with rank 1 keeps decreasing slowly and stably. The other two models shows wibbling but overall increasing out-of-sample error. 

\end{document}